# Unlocking the Power of Health Datasets and Registries: The Need for Urgent Institutional and National Ownership and Governance Regulations for Research Advancement

Ahmed S. BaHammam, MD

Department of Medicine, University Sleep Disorders Center, King Saud University, Riyadh, Saudi Arabia

**Running title:** Big Health Data Governance

**Funding:** No fund

**Corresponding author's full contact details:**

**Prof. Ahmed BaHammam**

ORCID: 0000-0002-1706-6167
Professor of Medicine
University Sleep Disorders Center, Department of Medicine, College of Medicine, King Saud University
Box 225503, Riyadh 11324, Saudi Arabia
Telephone: 966-11-467-9495
Fax: 966-11-467-9179
E-mails: ashammam2@gmail.com

**Abstract**

Health datasets have immense potential to drive research advancements and improve healthcare outcomes. However, realizing this potential requires careful consideration of governance and ownership frameworks. This article explores the importance of nurturing governance and ownership models that facilitate responsible and ethical use of health datasets for research purposes. We highlight the importance of adopting governance and ownership models that enable responsible and ethical utilization of health datasets and clinical data registries for research purposes. The article addresses the important local and international regulations related to the utilization of health data/medical records in research, and emphasizes the urgent need for developing clear institutional and national guidelines on data access, sharing, and utilization, ensuring transparency, privacy, and data protection. By establishing robust governance structures and fostering ownership among stakeholders, collaboration, innovation, and equitable access to health data can be promoted, ultimately unlocking its full power for transformative research and improving global health outcomes.

**Introduction**

"*Data is the new science. Big Data holds the answers.*"

–Pat Gelsinger

In the dynamic landscape of healthcare, we find ourselves navigating the vast terrain of big data—a realm brimming with immense potential to transform the way we understand, diagnose, and treat medical conditions. With the exponential growth of health data and the remarkable advancements in technology, we now live in an era where the power of data has the capacity to revolutionize healthcare delivery and improve patient outcomes based on real-world data [1, 2]. Unfortunately, the untapped potential of vast health data reserves represents a significant missed opportunity and a substantial waste of valuable resources.

With the expected launch of the Saudi Unified Health File, we expect to have substantial national data that will offer numerous benefits that significantly advance research and decision-making in healthcare [3]. This wealth of data facilitates more accurate and robust research studies, enhances evidence-based decision-making, and empowers healthcare professionals with valuable insights for improving patient care, optimizing healthcare delivery, and addressing public health challenges.

Medical records data encompass a wide range of patient information, including demographics, medical history, diagnoses, medications, treatments, diagnostic tests, imaging results, and outcomes such as deaths. This comprehensive follow-up enables a holistic assessment of healthcare services, including the identification of medical errors, adverse drug reactions, fraudulent activities, adherence to clinical guidelines, the effectiveness of treatments, optimal care pathways, and identification of individuals who respond well to specific treatments.

Nevertheless, despite the significant potential benefits and advantages offered by the era of big data, the machine-learning field has identified a range of ethical and legal issues. These concerns encompass various challenges, including the potential occurrence of representational harms, bias effects, breaches of privacy, unauthorized data access, inadequate governance, violations of local regulations regarding personal information, conflicts among stakeholders, and the presence of uncertain or questionable downstream applications [4-6]. In addition, preserving privacy and confidentiality within healthcare databases has raised concerns [7, 8].

Big data management in healthcare is complicated and requires transparent governance and regulations that adhere to International and National privacy and data protection regulations [9]. Amidst the transformative power of big data, it is crucial to address these challenges in order to navigate the ethical complexities and ensure responsible and equitable utilization of data-driven technologies.

In order to uphold patient privacy, several regulations have been implemented to regulate the utilization of clinical data by researchers. These include the Health Insurance Portability and Accountability Act of 1996 (HIPAA) in the United States (US) and the European Union General Data Protection Regulation [10, 11]. Additionally, institution-specific guidelines and governing bodies such as institutional review boards (IRBs) and Clinical Trial Units play a role in overseeing research involving patient data and other sensitive information found in electronic medical records, addressing concerns about the liability of healthcare providers and institutions [12, 13]. However, although ensuring access to clinical data is essential for driving translational research forward, there is always a concern among researchers that the multitude of regulations and policies governing the use of such data, while vital for safeguarding patient privacy and preventing misuse, frequently pose obstacles to data access and sharing.

Therefore, before we delve into the utilization of big health datasets in research, we must first explore the transformative possibilities and the governance and ethical considerations accompanying our existence in the big data era of healthcare. Consequently, it becomes imperative to establish robust institutional governance and well-crafted policies for medical records datasets that adhere to both national and international regulations, as they serve as the primary focal point for ensuring data security and compliance.

The central focus of this article is the importance of adopting governance and ownership models that enable responsible and ethical utilization of health datasets and clinical data registries for research purposes. It underscores the necessity of establishing explicit guidelines concerning data access, sharing, and utilization, with a strong emphasis on ensuring transparency, privacy, and data protection

**Data Ownership**

Various types of clinical data are frequently utilized in clinical and translational research. These encompass fully identified clinical data, HIPAA-limited clinical data, de-identified clinical data, and synthetic data [14]. HIPAA-limited clinical datasets consist of observational patient data that include restricted personally identifiable information (PHI), such as specific dates like admission, discharge, and service dates, as well as limited demographic details like city, state, zip codes, and age expressed in years, months, days, or hours. On the other hand, de-identified clinical datasets consist of observational patient data where all personally identifiable elements have been removed. In some regulations, accessing de-identified clinical data sets does not require IRB approval, although seeking an IRB Request for Determination of Human Subjects Research is advisable. These de-identified datasets can be used for clinical interpretation, scientific inference, and discovery, albeit to a lesser extent compared to HIPAA-limited clinical data sets. However, it should be noted that key variables or covariates may have been removed, limiting the ability to draw conclusions from de-identified datasets. For example, the absence of dates hinders the analysis of seasonal patterns in clinical outcomes and correlations with natural disasters, protocol changes, or regulatory issues such as new black-box warnings.

However, a fully executed Data Use Agreement (DUA) may also be necessary for identified patients' data, like electronic medical records. In order to fully empower full access and control over patient-centered health data, there is a crucial requirement for a data use DUA to be established. This agreement is executed between the patient and the entity responsible for providing healthcare and managing the patient's digital health record, known as the health data manager [15]. The DUA must encompass provisions that address important aspects such as data quality, integrity, privacy, security, and patient control. It needs to be centered around the patient's needs while also instilling trust and enabling clinicians and payers to utilize the patient's digital health data as necessary. Furthermore, DUAs can be customized to accommodate patient preferences, allowing them to contribute data for secondary purposes.

Collecting data proactively for research purposes, following the required ethical and institutional approvals, differs from the subsequent utilization of clinical patients' data originally processed within a medical healthcare facility for research, presenting a more complex scenario. Unfortunately, some researchers are unaware of this distinction, leading to conflicts between researchers and health research organizers within the institute. Therefore, the first step in adequately governing patients' medical and different health data is to define who owns medical information. Is it the healthcare provider, the patient receiving care, the one paying for it, or the doctor providing care? Or none of them or all of them? The question of ownership raises important considerations, but ultimately, does it hold significance? In fact, addressing this issue is of utmost importance and requires thorough resolution and organization prior to embarking on the utilization of large health datasets in research.

The concept of ownership establishes the practical ability or legal right to restrict others from using a particular thing [16]. Given this understanding of "ownership", it is not surprising that the notion of data ownership remains a topic of ongoing discussion and debate, despite the presence of comprehensive regulations in many developed countries [17]. While some advocate for the privatization of data (Privatization Postulate), viewing private ownership as a means to exert control over personal data, privacy, and property [18], others argue in favor of open science, considering individual-level health data as a common good (Communization Postulate) or public ownership [19].

Therefore, some propose shifting from "ownership" to "stewardship principles" or "good data sharing practices" [16, 20]. Nevertheless, this shift will not solve this issue, and institutional regulations need to define the responsibilities and privileges of each party.

The above expresses a part of the complexities related to the governance and ownership of health data, which also varies depending on different laws, regulations, and policies in different countries and institutions. Patients generally have some rights over their medical data, but the extent of those rights may vary. For example, through HIPAA Privacy Rule, patient data are protected, and patients have privacy and security around the information [10]. This means that patients must permit healthcare providers (i.e., hospitals and medical centers) to share their data with other healthcare organizations or researchers, which can be done through DUAs.

However, in most current regulations in developed countries, patients generally do not own their medical records (especially when de-identified), as they delegate that to the "*healthcare service providers*" upon opening health medical records and accepting to undergo management. For example, in 50 states in the US, medical providers typically own medical records, not patients [21]. Twenty-one states have statutes or regulations stating that providers own medical records [22, 23]. However, DNA and genetic material have generated their own body of distinctive law that is worth exploring in more detail. Though some states in the US have granted patients ownership rights to their genetic data not authorized for research purposes, judges have hesitated to acknowledge clear genetic ownership interests in other states [22].

Therefore, health data ownership poses complex challenges, influenced by various types of clinical data and regulations. Establishing DUAs is a way for enabling the utilization of patients' data in research while maintaining patient autonomy. The concept of data ownership remains a subject of ongoing debate, with perspectives ranging from privatization to open science and stewardship principles. However, institutional regulations must define the responsibilities and privileges of each party involved. While patients generally have rights over their medical data, the extent of those rights varies, and healthcare providers often hold ownership of medical records. Resolving these complexities and establishing transparent governance and ownership frameworks are vital prerequisites before leveraging large health datasets for research purposes.

**Regulations for Clinical Data Registries**

Healthcare providers are generally defined as organizations that provide medical or health services, including hospitals, clinics, skilled nursing facilities, and home health agencies. As healthcare providers are responsible for the collection, safeguarding, and accessibility of data, as well as being held accountable for any breach, leakage, or misuse of such data, they logically become the custodian of the data.

Health data obtained from patients' medical records are used in clinical research, longitudinal follow, and constituting clinical data registries. Clinical data registries are based on patients' data and medical records, which takes us to the initial question of who owns data to approve the development of clinical registries.

In the US, the ownership of clinical data registries is governed by state laws, which can vary depending on the state where the data originated, where it is stored, or where the primary offices of the Registry are situated [24]. In the majority of states, healthcare providers (hospitals) are granted ownership of the medical records they maintain from patient encounters [24].

Institutional regulations should address the role of hospitals in approving and supervising clinical registries and their usage to ensure the confidentiality of patient data. It is essential to obtain the approval of the service provider (hospital) when developing registries to maintain the confidentiality code entrusted to hospitals and prevent unauthorized access to non-deidentified data by unauthorized personnel. Failing to obtain hospital approval violates the agreement between the hospital and patients and breaches the confidentiality code. Therefore, it is crucial to establish proper protocols and oversight mechanisms to safeguard patient data and maintain the integrity of clinical registries.

To eliminate any ambiguity or potential disputes in clinical data registries, the regulations should explicitly outline the legal distinctions in relation to data ownership. The regulations should clearly address: (i) patients' rights, (ii) the service provides' rights, (iii) upholding the protection of the Participant's data in accordance with HIPAA and other relevant laws for as long as the Registry possesses the data, (iv) roles and responsibilities of the sponsor, (v) identifying stakeholders at an early stage of the registry planning process, (vi) outlining the procedures through which registry investigators obtain access to and analyze registry data to create abstracts for scientific conferences and draft manuscripts for submission to peer-reviewed journals, and (vii) the process of sharing data, as outlined in the Data Sharing Regulation released by the National Data Management Office [25] and other international regulations [26, 27].

Furthermore, it is crucial to proactively identify stakeholders during the initial phases of establishing a registry. To determine the relevant stakeholders, it is important to consider who is directly impacted by the research questions. Identifying these stakeholders early in the planning process allows for valuable contributions in shaping the nature and extent of data collection. Additionally, these stakeholders may serve as end-users of the data or play a significant role in disseminating the registry's findings. Their involvement at an early stage can greatly enhance the registry's effectiveness, success, and impact.

Registries also need meticulous governance and supervision to ensure the registry's responsible and efficient operation, uphold scientific integrity, and comply with ethical and legal requirements. It is proposed that the functions of governance and oversight in a registry can be classified into the following categories: executive or steering function, scientific function, liaison function, adjudication function, external review, and data access, use, and publications function (Figure 1) [28].

- The executive or steering function is responsible for making critical decisions regarding the financial, administrative, legal/ethical, and scientific aspects of the registry. This function requires input from experts in relevant domains to ensure informed decision-making. In certain situations, the group responsible for the steering function may also fulfill additional roles described subsequently.

- The scientific function comprises experts from diverse fields, including database content, clinical research, epidemiology, and biostatistics. They play a key role in guiding the registry's inquiries and advising the executive or steering group on specific analyses. It is essential that the registry's reports uphold high scientific standards, ensuring rigor, independence, and transparency. To enhance credibility, stakeholders' roles in the publication process should be clearly defined, and any potential conflicts of interest must be disclosed.
- The liaison function is important in larger registries and involves establishing and maintaining relationships with the funding source, healthcare providers, and individuals seeking registry information. The group responsible for this function may develop tools to monitor and assess user satisfaction, ensuring the registry operates efficiently on a daily basis.
- The adjudication function involves reviewing and confirming challenging cases or outcomes without knowledge of the exposure being studied, ensuring unbiased confirmation.
- External review committees or advisory boards provide independent oversight throughout the registry's duration, with periodic reviews for safety data accumulation in certain situations.
- Regarding data access, use, and publications, the function addresses how registry investigators access and analyze data for scientific presentations and publications. Authorship guidelines, such as those outlined by the Uniform Requirements for Manuscripts Submitted to Biomedical Journals, should be followed, considering potential variations based on funding sources or biomedical journals. Furthermore, there should be a review process to evaluate and respond to requests from other researchers or entities seeking access to the data for specific purposes, particularly in registries generating external interest.

**Regulations and Governance of Big Health Datasets in Hospitals and Healthcare Facilities**

Before creating big clinical data repository for research purposes, it is essential to establish policies that restore patient agency and autonomy and involve all stakeholders. In addition, international and national regulations recommend setting policies and regulations for data classification, sharing, privacy, freedom of information, open data, and others [9, 29].

The National Data Management Office (NDMO) in Saudi Arabia has mandated the development of policies, governance mechanisms, standards, and controls related to data [25]. Therefore, before establishing a clinical data hub based on medical records for research in any institute, the first and most important step is for the institute to develop transparent governance and standard operating procedures for patients' health data. The NDMO proposed eight governance domains and 28 specifications and data controller's privacy policies and procedures [25]. The domain grants the jurisdiction and power to govern and execute the organization's data management procedures. The NDMO encompassed health data, as well as images or videos related to individuals, within its data regulations. Furthermore, it emphasized the importance of identifying, documenting, and obtaining approval from the head of the entity (or their representative) for the privacy policies and procedures of the Data Controller. These policies should be shared with all relevant parties involved.

Additionally, the responsibility lies with the Data Controller to create and implement policies and procedures related to the protection of personal data. The approval of these policies and procedures rests with the head of the entity or their representative [30]. Following this, the Data Controller should establish an organizational unit, connected to the Chief Data Officer (CDO) Offices in government agencies as established by Royal Decree No. 59766 on 20/11/1439 AH. This unit will be tasked with developing, documenting, and overseeing the implementation of the policies and procedures. The NMDO regulation also covers who is the owner of Analytic Data Assets [30].

The regulation also covered who is the owner of Analytic Data Assets [30]. They indicated that the initial step is to formally constitute a Data Governance Steering Committee that includes all stakeholders. The committee should develop the executive policies, procedures, standards, ownership, decision rights, roles and responsibilities, and accountability related to the data while abiding by the national regulation. Furthermore, a governance model that recognizes clinicians and patients as key stakeholders and includes them as members is recommended in most international regulations.

Moreover, to ensure effective utilization of health data for research purposes, it is important to implement several key measures as endorsed by the World Health Organization [31]. Firstly, the establishment of a dedicated information technology (IT) infrastructure specifically designed for research activities should be prioritized, keeping it separate from operational data processing activities. This segregation helps maintain the integrity and security of both datasets. Furthermore, the active engagement of researchers with the available data and providing them with a comprehensive education on big health data research privileges are crucial. Furthermore, developed institutional policies should cover the rights and responsibilities of researchers, and researchers should be well-informed about their rights and responsibilities when handling and analyzing data to maximize its potential for research purposes. Conducting cross-border research requires a meticulous assessment of the legal framework in place. Therefore, it is essential to thoroughly understand the regulations and requirements governing data sharing and privacy across different jurisdictions. This ensures compliance with the relevant laws and safeguards the rights of individuals involved in the research. Developing policies to define potential data applications and preventing the occurrence of data-related harm to patients, such as breaches of privacy, discrimination, stigma, disenfranchisement, disempowerment, and exploitation, where individuals or groups may face disadvantages due to the use of their data to characterize them [32]. Lastly, maintaining proper documentation of research activities is of paramount importance. This includes documenting justifications for the secondary use of data and retaining data for research purposes. By documenting research activities thoroughly, researchers can ensure transparency, facilitate future analysis, and validate research findings.

Collectively, implementing these measures promotes the responsible and effective utilization of health data for research, fostering advancements in the field and contributing to improved healthcare outcomes. Finally, The institute and controllers should ensure that data protection should not be a barrier to research.

**Other Barriers to the Establishment of Big Health Datasets in Hospitals and Healthcare Facilities**

Besides governance, implementing big clinical data for research purposes in hospital settings has several obstacles [33]. One of the main challenges is ensuring data accuracy, completeness, and consistency, which is crucial for reliable analysis and decision-making. Data entry errors, missing data, and inconsistencies can compromise the quality of big clinical data. Integrating data from various sources and different electronic health records systems, such as medical imaging and laboratory results, can also be difficult due to differences in data formats, standards, and terminologies. Therefore, interoperability between systems is essential for seamless data exchange and analysis.

Another challenge is ensuring compliance with regulations like HIPAA and GDPR (General Data Protection Regulation) while safeguarding confidential and sensitive information from unauthorized access and potential breaches. Hospitals must invest in infrastructure and technologies that can handle the storage, retrieval, and processing of large datasets.

Moreover, analyzing big clinical data requires advanced analytical tools and techniques like machine learning and artificial intelligence. Therefore, hospitals and academic centers need to invest in skilled personnel and resources to exploit the full potential of big clinical data.

Addressing these barriers will require collaboration between administrators, policymakers, funding agencies, and researchers to develop solutions that promote the proper utilization of big data while also protecting patient privacy and confidentiality.

**Summary and Conclusions**

Effective governance of big health data and clinical data registries can enhance health information systems, optimize data utilization for improved quality, safety, and performance outcomes, and facilitate advancements in preventive practices and medical diagnostics, treatments, and practices. However, achieving effective health data governance necessitates striking a balance between the interests of various stakeholders, including data subjects, researchers, sponsors, and society. Additionally, it requires overcoming a range of challenges encompassing technological, economic, legal, and ethical aspects. To establish consistent, sustainable, well-regulated, and low-vulnerability institutional work, it is essential to start by developing institutional regulations for medical records datasets that abide by national and international regulations.

Moreover, to foster responsible use of health data, it is crucial for all stakeholders involved in the health data cycle to comprehend their obligations and entitlements and establish trust-based models that enable ethical data sharing, focusing on protecting individual identities and promoting collective and participatory decision-making processes.

Additionally, though several organizations and papers use the term "ownership" for patients' data, it is better to shift the focus toward examining the rights and regulations that individuals, groups, and administrations have regarding data. It should also consider both patients and societal perspectives. This comprehensive, systematic, and collaborative approach to big health data will not only help us avoid potential difficulties down the line, but also promote a sense of shared responsibility and ownership, which is crucial for the success of any project.

Finally, this article seeks to create a foundation for establishing ethical guidelines, defining ownership structures, and implementing transparent governance practices for health datasets and clinical data registries. Its goal is to promote the institutional and national responsible and fair use of data-driven technologies. Without the implementation of institutional and national strategies that prioritize the safe utilization of health data and the establishment of a robust health data governance framework, numerous institutions, organizations, and countries may fail to seize the potential of data use in enhancing healthcare quality and performance in a secure manner.

**Figures Legends**

**Figure 1:**

An outline of the proposed clinical data registry governance functions to ensure the registry's responsible and efficient operation, uphold scientific integrity, and comply with ethical and legal requirements.

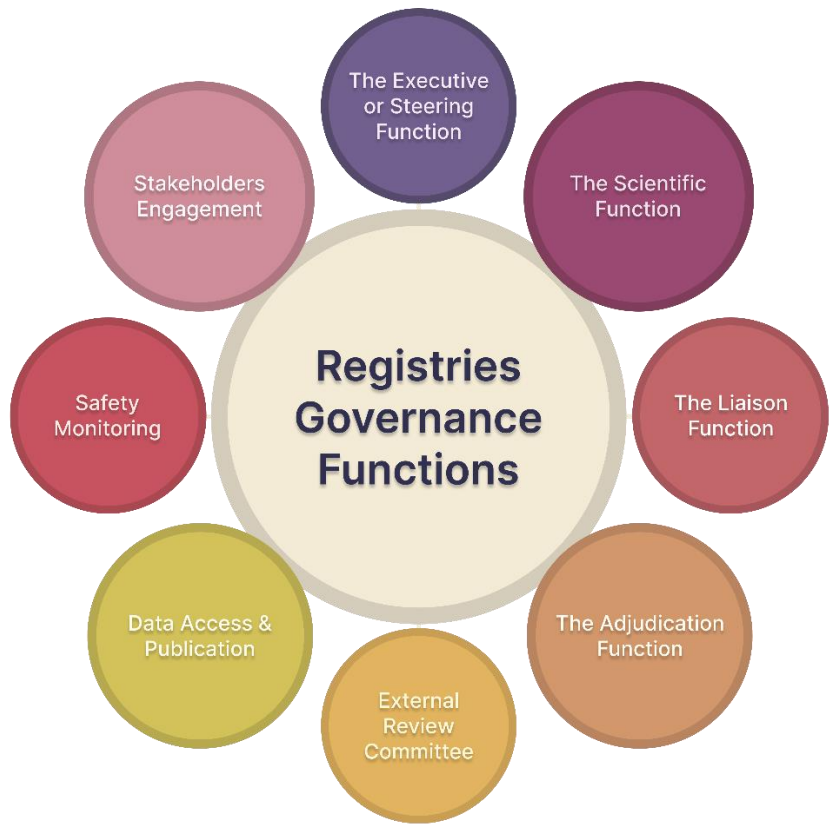